# Monte Carlo calculation of organ and effective dose rates from ground contaminated by Am-241: Results of an international intercomparison exercise

J Eakins[1,*], C Huet[2], H Brkić[3], K Capello[4], L Desorgher[5], L Epstein[6], J G Hunt[7], H S Kim[8], D Krstic[9], Y-K Lee[10], M Manohari[11], D Nikezic[9,12], R H Shukrun[6], D. Souza-Santos[7], K. Tymińska[13]

[1] *Public Health England (PHE) CRCE, Didcot, Oxfordshire, United Kingdom*
[2] *Institute for Radiological Protection and Nuclear Safety (IRSN), Fontenay-aux-Roses, Paris, France*
[3] *Faculty of medicine in Osijek, Osijek, Croatia*
[4] *Health Canada, Ottawa, Ontario, Canada*
[5] *Institute of Radiation Physics (IRA), Lausanne University Hospital and University of Lausanne, Lausanne, Switzerland*
[6] *Soreq Nuclear Research Center, Yavne, Israel*
[7] *Instituto de Radioproteçao e Dosimetria, Rio de Janeiro, Brasil*
[8] *Korea Institute of Radiological and Medical Sciences (KIRAMS), Seoul, Republic of Korea*
[9] *University of Kragujevac, Faculty of Science, Kragujevac, Serbia*
[10] *Université Paris-Saclay, CEA, Service d'Études des Réacteurs et de Mathématiques Appliquées, 91191, Gif-sur-Yvette, France*
[11] *Indira Gandhi Centre for Atomic Research, Kalpakkam, Tamil Nadu, India*
[12] *State University of Novi Pazar, Novi Pazar, Serbia*
[13] *National Centre for Nuclear Research, Otwock, Poland*

* Contact: jonathan.eakins@phe.gov.uk

ABSTRACT

An intercomparison exercise is described that examines Monte Carlo modelling of anthropomorphic voxel phantoms in an idealized ground-contamination photon exposure scenario. Thirteen participants calculated and submitted organ and effective dose rates for comparison against a set of verified reference solutions. The effective dose rates are shown to agree with the reference value to within reasonable statistical uncertainties in five of the cases, though in only one of those was similar agreement also demonstrated in the evaluation of all requested organ dose rates. Orders-of-magnitude differences in doses are seen for some of the other participants, both internally within their own dataset and also relative to the reference solutions. Following limited feedback and suggestions from the organizer, up to two sets of revised solutions were resubmitted by some of the participants; these generally exhibited improved agreement, though not always. The overall observations and conclusions from this intercomparison exercise are summarized and discussed.

INTRODUCTION

Working Group 6 (WG6) of the European Radiation Dosimetry Group (EURADOS) recently organized an intercomparison study [Zankl *et al*, 2021a] on individuals' uses of radiation transport codes with the International Commission on Radiological Protection (ICRP) reference computational voxel phantoms (RCP) [ICRP, 2009]. Various exercises were defined that required participants to evaluate specific dose quantities, and were intended to be of practical interest in occupational, environmental or medical dosimetry. The aim was for participants to attempt the tasks and submit results that could be compared against reference solutions derived by the organizers. The overall purposes of the endeavour were to: investigate how well the phantoms have been implemented by the participants in their models; allow participants to check their calculations against quality-assured master solutions; provide an opportunity for participants to improve their approach via feedback; and identify common pitfalls and difficulties that could serve as general lessons learnt. In addition, extrapolation of the findings was hoped to give some insight into the general status of voxel phantom usage within the wider computational dosimetry community.

The current paper focusses on just one of the exercises, which related to a scenario in which a person was standing on ground contaminated by a photon-emitting radionuclide. A summary of the configuration that was to be modelled is provided first, followed by a presentation of the results that were initially volunteered by the participants. These data are then augmented by resubmitted results, which were provided subsequently by the participants following limited feedback from the problem organizer. Finally, discussion is given on the general trends exhibited in the submissions, the common successes and mistakes made by the participants, and the overall conclusions from the intercomparison exercise.

PROBLEM SET-UP

The overall goal of the exercise was to calculate organ absorbed dose rates for the adult male (RCP-AM) and adult female (RCP-AF) reference computational phantoms, as well as the effective dose rate, from an idealized photon exposure scenario representing ground contaminated by Am-241. The problem was specified according to the following description, which matches the information provided to the participants:

- The ground was concrete of depth 0.5 m, density 2.3 g cm$^{-3}$, and composition as defined in Table 1.

- For simplicity, the Am-241 was to be approximated as a monoenergetic source of 60 keV photons.

- The contamination was assumed to be contained within a disc of radius 2 m, with the anthropomorphic phantom standing at its centre, and was deposited on the surface of the ground only. The photons were emitted isotropically ($4\pi$ solid angle) from this planar surface.

- A uniform ground contamination was assumed, with an emission rate of 10$^6$ photons cm$^{-2}$ s$^{-1}$.

- The entire configuration was surrounded by vacuum, and is illustrated in Figure 1.

It was recommended that participants use the reference computational phantoms as described in ICRP Publication 110 [ICRP, 2009], with the organ and tissue masses that are given therein. For red bone marrow and endosteum (bone surface) dosimetry, the method proposed in ICRP Publication 116 [ICRP, 2010] was recommended: that is, application of dose response functions or dose enhancement factors. For the calculation of effective doses, the tissue weighting and radiation weighting factors from ICRP Publication 103 [ICRP, 2007] were presumed, though the latter equals 1 for photons.

Participants were tasked with reporting the organ absorbed dose rates from the contamination to the brain, lungs, small intestine (SI), stomach and red bone marrow (RBM) of both the RCP-AM and RCP-AF, as well as the overall effective dose rate. Template spreadsheets were provided to participants, in which they could enter their solutions in a pre-defined format to facilitate the evaluation process. The template also asked participants to document the transport code that was used, whether the kerma approximation was adopted, and which photon and electron cross-section libraries were implemented, as well as basic personal details regarding their country of origin and institute etc. Participants were also requested to state explicitly the method of bone dosimetry that they used, and asked to explain in detail any method that deviated from that of ICRP Publication 116.

Two members of EURADOS WG6 coordinated the exercise, an 'organizer' and a 'co-organizer'. Prior to opening the intercomparison, the task was first completed by the organizer to generate a master solution. It was also then independently completed by the co-organizer, to ascertain and confirm the correctness of that master dataset. With agreement established within acceptable statistical precision (< few %), this solution was used as a reference in the subsequent analyses of the participants' submissions. In those analyses, a given participant's result was assumed to be in agreement with the reference solution if the two values were proximal to within their combined statistical uncertainties ($k$=1). Whilst clearly this is not a statistically rigorous definition, it was assumed adequate for the subsequent qualitative discussions on whether or not the participants had been broadly successful in their attempts of the exercise.

INITIAL RESULTS

In total, thirteen participants submitted solutions for this ground contamination problem. These are identified by randomly allocated numbers and letters in the following, with all results presented anonymously. The intercomparison was truly international: participants originated from ten different countries, one from Canada, Croatia, France, Israel, Poland, Serbia, South Korea, Switzerland and Vietnam, and two each from Brazil and India. The computer codes, particle transport options, cross-section data and RBM dosimetry methods that were reported by the thirteen participants (labelled 1 to 13) are summarized in Table 2, along with the analogous details used in the generation of the reference solutions by the organizer.

Participant 13 provided no details on how the calculations were performed or what code or data they had used. Although deterministic codes could conceivably have been employed, all of the remaining twelve participants used Monte Carlo software. It is clear from Table 2 that the MCNP family of codes was the most widely adopted, with six participants using one of those versions. Participant 6 submitted two solutions for the male phantom, one of which used the kerma approximation whilst the other employed full coupled electron-photon transport; just the kerma approximation was used by that participant for the female phantom. However, their results from the two methods were statistically irresolvable, so for convenience only the former are shown and discussed in the following. Four of the other participants (3, 4, 5 and 7) made the kerma approximation, as did the organizer for the generation of the reference solutions. Generally, the participants used the latest cross-section libraries that were available to them. Half of the participants stated that they employed the ICRP 116 method for estimating RBM doses; based on what they reported, it is possible that Participants 4, 5, 8, 10 and 11 also used this scheme, but insufficient details were provided to verify this. The 'homemade' method employed by Participant 12 for RBM dosimetry was not described further.

The organ and effective dose rate data that were generated by the organizer are presented in Table 3. Also shown are the concurrent standard uncertainties that relate just to the statistical fluctuations within the Monte Carlo code. The standard uncertainty of around 0.1 % for the effective dose rate was calculated in quadrature from the uncertainties on the twenty-nine weighted organ doses used in its definition [ICRP, 2007]. The organ dose rate results that were submitted by the thirteen participants are shown in Figure 2 for the male phantom and in Figure 3 for the female phantom, along also with the reference results. Note that the previous numeric participant identifiers are now replaced by new randomly allocated letters (A to M) in Figures 2 and 3, to preserve anonymity and remove any apparent connections between, for example, performance and code usage; to that end, no correlations exist between the letters and numbers, so Participant A ≠ Participant 1 etc. In both figures, the top left plot shows all of the data that were submitted, the top right plot shows that same data on a partially restricted *x*-axis that serves to remove the most extreme outliers, and the bottom plot shows the data on a more severely restricted *x*-axis that allows greater resolution of the majority of the results; the exact ranges used in the latter two cases were chosen somewhat arbitrarily.

Participant L submitted a result for the male RBM dose rate that was about 8 orders of magnitude too high. Their value for the female RBM was consistent with their other organ dose rates, however, as was their estimate of effective dose rate. This suggested an oversight by that participant, for example by neglecting to apply the correct normalization or including a typographical mistake in their submission, rather than a genuinely severe fault in their modelling. Indeed, subsequent correspondence with the participant confirmed this hypothesis, with the individual reporting that they had in fact calculated a dose rate of $5.845 \times 10^{-8}$ Gy/s for the male RBM rather than the quoted value of 5.845 Gy/s. Such a corrected result would still be around 60 % higher than the reference solution, however, and is comparable to the over-estimate (about 50 %) also seen in the female RBM dose rate from that participant. For Participant G, both the male and female RBM dose rate results were around an order of magnitude too high, but in fact all results from that participant were much higher than the reference solutions. This might suggest a genuine error in their model, or misunderstanding of the correct evaluation of organ doses and RBM dosimetry.

Even when the most obvious outliers are excluded, there is still a very wide range of results from the participants. In general, a broadly similar spread is exhibited in the male and female datasets, indicating as expected that neither phantom is any more 'troublesome' to use than the other. Of all the organs, the RBM caused the most discrepancy, both in terms of general magnitude and frequency: only Participants F and J provided male and female results that agreed with the reference solutions within uncertainties, though Participant M also came close (agreement within 5 %). To put that in context, four out of the thirteen participants (C, I, J, M) provided results that agreed within uncertainties for each of the brain, lungs, small intestine and stomach of both the male and female phantoms; moreover, in some of the remaining eight cases (B, D, F), agreement was observed for one of the phantoms but not the other, whereas such partial success was not exhibited in any of the RBM results. Given the more complex dosimetry of the RBM compared to the other organs [ICRP, 2010], these observations are perhaps unsurprising.

The effective dose rate results, $E$(P), that were submitted by the participants are shown in Figure 4 as ratios, [$E$(P)/$E$(O)], to the reference effective dose rate, $E$(O). Participant H did not calculate effective dose *per se*: rather than the correct sex-averaged quantity, they instead provided solutions for the male and female phantoms individually. These are shown separately on Figure 4, noting that the male result of 4.66 and the female result of 4.77 would have given a ratio of 4.72 had averaging been applied. It was observed also that Participant G calculated effective dose by averaging just the data for the five organs requested in this exercise, rather than weighting and summing all twenty-nine organs specified in its definition [ICRP, 2007], so is also not an accurate estimate of the correct dose quantity; this participant also exhibited the largest divergence from the reference value.

Five participants (D, F, I, J and M) submitted solutions for effective dose rate that agreed with the reference value to within reasonable statistical variations, with a further three (B, C and L) agreeing to within about 10%. This is perhaps surprising, because only Participant J agreed with the reference results for all ten of the reported organ dose rates (Figure 2 and 3). The explanation of why greater success was generally apparent for effective dose compared to organ doses is possibly that the weighting / averaging process served to mitigate individual organ results that were too high or too low. This may be the case especially for the RBM, which was often poorly evaluated: when tissue-weighted by 0.12 and aggregated with the other organs, the impacts of this outlier would have lessened. The implication from this is that only Participant J determined effective dose correctly, whilst the statistical uncertainties on the results of Participants D, F, J and M hid the underlying systematic errors. Of the five solutions (A, E, G, H and K) where large discrepancies are evident in [$E$(P)/$E$(O)] (Figure 4), it is interesting to note that it was always because the participant solution was substantially larger than the reference value, rather than the other way around. It is not clear whether this apparent pattern is simply a coincidence, or if not, what its causes and significance might be.

Statistical uncertainties were reported by the participants for all their results, and arose from the inherently stochastic Monte Carlo method of solution. However, these have not been included in Figures 2-4 for clarity. Instead, the relative standard uncertainties on the organ dose rates are shown in Figure 5, expressed as a percentage. Of note are the differences in these uncertainty data from one participant to another, from one phantom to the other for a given participant (e.g. L), and between the various organs of a given phantom for a given participant (e.g. C, D, G). Some of the participants (e.g. C, D, G and L) reported uncertainties that were consistently higher than those of the other participants. Running greater numbers of particle histories could presumably have helped in these circumstances, though this may not have been a practical solution for everyone given the concurrent increases in CPU times required to perform them: typically, decreasing uncertainties obey a Poisson relationship, so scale with the square of the simulation time. Prior to performing the re-anonymization process, the organizers observed that, perhaps surprisingly, no obvious correlation existed between lower statistical uncertainties (Figure 5) and use of the kerma approximation by the participants (Table 2). Photon-only transport is expected to be computationally more efficient than full coupled electron-photon transport (see e.g. [Werner *et al.*, 2018]), so presumably those participants who employed the latter required longer CPU times to arrive at the same degree of precision. It is noted also that some of the uncertainty datasets contained obvious outliers, an example being the stomach of the male phantom for Participant L. There is no obvious reason why such a large organ would result in such a comparatively high uncertainty, especially considering that the dose to which it corresponded agreed with the reference value to <1 %, and also that an uncertainty of <1 % was reported by that participant for the dose to the stomach of their female phantom.

The relative standard uncertainties quoted for the effective dose rates were typically small. The organizer reported a value of 0.1 % (Table 3), and all participants reported values less than 1 % apart from Participants C (5 %), D (3 %), G (19 %) and L (2 %). All ten organ doses were reported by Participant D as having uncertainties of either 3 or 5 %, without further precision, and the uncertainty quoted for the effective dose was perhaps surprising given that the quantity is an aggregate. However, the weighted doses and uncertainties for the many other organs used to determine effective dose [ICRP, 2007] were not provided by (or requested from) the participants, so such an analysis is hard to conclude. As mentioned previously, the method applied by Participant G to calculate effective dose was incorrect, so interpretation of the uncertainty estimate reported alongside it is similarly limited.

REVISION OF RESULTS

Following initial analyses and presentation of the results, the organizer contacted the thirteen participants separately to provide bespoke feedback. For one participant (J), this feedback was simply an acknowledgment of their successful agreement with the reference solution; for the remaining twelve, limited information was provided that highlighted the main areas of disagreement. This information was kept deliberately vague: the intention was to notify the participants of which results diverged from the reference values and by approximately how much, with the aim that they could then use that insight to inform self-analyses of their modelling. The participants were also invited to resubmit a revised set of results that, hopefully, would better-agree with the reference solutions. To illustrate this approach, example feedback was of the form:

- *You agreed well with the reference solutions apart from the RBM, both results for which were too high by a factor of a few;*

- *Most of your results were higher than the reference solutions by a factor of around 2-3, whilst your small intestine doses for both male and female were lower by a similar magnitude;*

- *For the female lungs you were about 8% lower than the reference value, which seems anomalous given the good agreement elsewhere and low uncertainties quoted;*

- *All your results were higher than the reference solutions by about an order of magnitude. You also provided two results for effective dose.*

Of the twelve participants who were invited to resubmit solutions, ten elected to do so. These revised results are shown in Figure 6, with the effective dose data again given relative to the reference value (i.e. [$E$(P)/$E$(O)]). Restricted ranges have been applied to Figure 6 for clarity. Specifically, full data for Participant A are excluded because they provided dose rates for all organs that were around 3 orders of magnitude too high, leading to an effective dose rate ratio of about 1400. This was surprising, as the feedback to that individual on the original submission was that most of their results only diverged from the reference values by a few tens of percent. Also difficult to resolve in Figure 6 are the RBM dose rates reported by Participants G and K, which were respectively three and eight orders of magnitude lower than both the reference solutions and the other organ dose rates in their datasets; possibly this indicates problems during renormalization, or similar error. As previously, Participant H did not provide a sex-averaged estimate of effective dose, instead submitting separate data for the male and female phantoms.

Where differences persisted in the participants' revised solutions, the individuals were contacted a second time to provide further feedback. Although still not prescriptive, this additional information was intended to be more specific than previously, in order to better help them identify and resolve their problems. Following their investigations, the participants were also invited to resubmit a third set of results, which was to be considered final. Examples of this subsidiary feedback include:

- *Your results are now 2 orders of magnitude too high. Perhaps there is a problem in your normalization, or in your conversion of *f8 energy deposition to dose rate?*

- *Your RBM results are now closer to the reference values, but still too high. The method I used for dose enhancement follows the recommendations of ICRP 116 with the data contained in that document (Appendix D). What is your method?*

- *Your results are still twice the reference values. Could this be a normalization issue or a problem with the source, for example?*

- *Effective dose is not calculated by taking the weighted average of just these organs: many other organs of the body need also to be included. You may find reference to the definition in ICRP 103 of use here.*

Of the eight participants who were invited to resubmit solutions a second time, seven elected to do so. Their revised results are shown in Figure 7, with the effective dose data again given as ratios to the reference value. Restricted ranges have been applied to Figure 7 for clarity. Specifically, full data for

Participant A are again excluded because they provided dose rates for all organs that were around an order of magnitude too high, leading to an effective dose rate ratio of about 11. In general, better agreement is exhibited in Figure 7 compared to Figure 6, though some discrepancies with the reference data persist, especially for the RBM dosimetry.

Some of the participants included comments or explanations of what changes they had made when they resubmitted a solution. This self-identification of errors is encouraging, as well as insightful into common causes of difficulties. These explanations may be summarized as follows:

- Three participants reported an error in their source definition, and/or in the correct normalization of their results to it. One participant initially used the full Am-241 decay as the energy distribution of their photon source; although not incorrect *per se*, this was wrong from the perspective of the intercomparison, for which the approximation of a 60 keV monoenergetic source was specified.

- Two participants reported mistakes in their methods used to calculate RBM doses. One of those subsequently reported that they were not using the fluence-to-dose response functions given in ICRP 116.

- One participant reported a geometry error, specifically the omission of the concrete floor.

- One participant reported using the wrong tally, specifically a pulse-height tally (MCNP *f8*) rather than an energy deposition tally (MCNP *\*f8*).

- One participant reported using incorrect photoelectric mass attenuation coefficients for one organ, presumably leading to erroneous estimates of energy deposition.

- One participant reported that they had made a mistake in copying and pasting their results.

- One participant simply reported that they had 'found some bugs and removed them', with their nature not specified further.

From the above comments, it is clear that the most common self-identified cause of error was incorrect definition of the source. Mistakes in RBM dosimetry were also identified, as expected, though these were not reported frequently enough to reflect the general lack of agreement exhibited by most participants for that organ. It is also noteworthy that with the exception of the RBM cases, which might conceivably have resulted from misunderstanding of the recommended dosimetry techniques, all of the reported mistakes were perhaps avoidable. Of course, this is a somewhat biased observation, because inevitably it is easier to self-identify 'typographical' or simple geometry errors than conceptual ones, but does emphasize the need for thorough checking and quality assurance of all input files and results.

Note that some participants sent amended results intended to update / replace data they had recently provided; these revised datasets were typically sent soon after that submission, and were accompanied by an explanation stating that the individual had noticed mistakes in their earlier set. In these cases, only the updated data have been included in the intercomparison exercise (Figures 2-7). This discretion has been permitted on the grounds that the participant had identified their mistake themselves during their own routine quality assurance (QA), without any prompting or additional feedback from the organizer, and that in many real-World dose assessment scenarios, individuals are typically able to review and revise their own results for errors even at a comparatively late stage.

SUMMARY AND DISCUSSION

In the current intercomparison exercise, a given result was taken to agree with the reference solution if the two values were proximal to within their combined statistical uncertainties ($k$=1); this provided a simple, qualitative indication of whether participants had been broadly successful in their attempts. Of course, it is acknowledged that a more statistically rigorous analysis may have been devised, and it remains an open question to be addressed within WG6 as to how this might best be achieved in general within intercomparison exercises (cf. [Zankl *et al*, 2021a]). Nevertheless, even with the heuristic approach adopted here, a number of observations and comments may be summarized from the analyses of the solutions provided by the thirteen participants (Figures 2-7):

- Only one participant provided an initial set of results that agreed with the reference solutions exactly (within statistics). Five participants provided solutions that were within a few tens of percent of the reference solutions; and three participants provided results that were within a few tens of percent of the reference solutions in most cases, but with a few organ doses deviating by much more.

- The remaining four participants provided initial solutions that were at least an order of magnitude different from the reference results. In one of those cases, the doses were all too high by a broadly consistent amount; in another one of the cases, the doses were also too high but by amounts that varied within the dataset; and in the last two cases, mixtures of large positive and negative differences were found within each dataset. In the latter cases, the results could have been rejected directly as being physically implausible.

- Overall, the RBM was the organ for which the participants' results differed most frequently from the reference solution: only two participants provided initial results that agreed within uncertainties for both phantoms, though one further participant agreed to within 5 %.

- Five participants submitted initial solutions for effective dose rate that were in acceptable statistical agreement with the reference value, although only one of those had used consistent values for RBM doses in its derivation. One participant calculated effective dose by averaging just the five organ doses that were requested specifically for this exercise, though tissue-weighting and sex-averaging were correctly applied. Another participant provided tissue-weighted equivalent doses separately for the male and female phantoms, with no further sex-averaging, so their results were only 'effective dose-like' rather than the correct dose quantity.

- All participants were contacted to provide feedback and 'hints': in the first instance this was kept deliberately vague, with more specific instruction then given a second time, if necessary. In some cases, participants found and explained the cause of their error(s), and their resubmission exhibited greater agreement with the reference solutions. However, some of the resubmissions were worse. Not all participants provided a resubmission, and not all discrepancies were able to be explained or accounted for within the lifetime of the intercomparison exercise

- When the three iterations of solutions are considered together, a total of three out of the thirteen participants arrived at final results for all organ dose rates that were statistically irresolvable from the reference value, and nine out of thirteen for the effective dose rate. Greater apparent success for the latter is likely due to the 'smearing-out' of individual outlier organ doses.

Participants were not required in the intercomparison exercise to submit explicit results for doses to the endosteal tissue within the bone surfaces. However, it may be reasonable to speculate that their success rates would have been analogous to those for RBM, because ICRP recommend similar calculation methods for both organs [ICRP, 2010]. Endosteal tissue is one of the organs that contributes to effective dose [ICRP, 2007], so in some cases this component might have contained inaccuracies that were similar to those for RBM, although their overall impact would have been less: the tissue weighting factor for endosteum is only 0.01, compared to 0.12 for RBM. This reinforces the suggestion that the greater success of the effective dose results compared to the organ doses may be somewhat illusory, and if more statistical precision were obtained, the divergence of the estimate from the 'true' organ-weighted value would become manifest. Nevertheless, it is an interesting observation: just because a modeller has arrived at a plausible-looking value for *E*, it does not imply that the concurrent organ doses associated with it are correctly evaluated. This would be an important point to consider in a real exposure situation if other doses were of interest but only effective dose had been used for benchmarking, for example by comparison against any measured data for $H_p(10)$ that might be available.

A number of participants submitted some results that were obviously wrong, such as a dose for one of the organs that differed greatly (order(s) of magnitude) from the rest of their dataset, which immediately may be flagged as unrealistic from a consideration of the underlying physics. Such occurrences perhaps indicate an oversight or lack of quality assurance on those occasions, and in some cases it may be fair to remark that this type of internal inconsistency should perhaps have been picked-up and investigated by the participant prior to submission. Either way, it is noted that those individuals often reported uncertainties on their data that were typically small, as in fact did most participants in general (Figure 5). The observation here must therefore be that low statistical uncertainties from Monte Carlo simulations can bely large systematic uncertainties, and are by no means an indicator of overall accuracy. Indeed, the anecdotal experience of the present author is that this point is not recognized nearly enough in the wider field of computational dosimetry: it is often obvious that very high levels of statistical precision have been strived for within a Monte Carlo simulation, with only those uncertainties then reported and discussed, and without full consideration of the much larger 'hidden' uncertainties intrinsic to the modelling.

Even in those cases where self-consistency is evident within a given dataset, there are still simple tests that may be applied to check its potential veracity. For example, one could easily add a small air-filled volume just in front of the phantom at around chest-level, and use it to tally the photon fluence at that height; on doing this for the male, the organizer obtained an approximate value of $2.2 \times 10^{-6}$ cm$^{-2}$ per-source-photon at that location. One could then use this value to normalize, say, the absorbed dose to the lungs that was calculated (about $3.0 \times 10^{-19}$ Gy per-source-photon, for the organizer), to give an approximate dose per fluence for that organ ($1.4 \times 10^{-13}$ Gy cm$^2$). Finally, one could compare this against the data given in ICRP Report 116 [ICRP, 2010] for the male lungs from, for instance, isotropic (ISO) or rotational (ROT) exposures to 60 keV photons. Of course, an ISO or ROT source in vacuum is different from that of ground contamination on a concrete floor (inferior-hemisphere semi-isotropic (IS-ISO) would be better), with shielding and backscatter from the phantom also perturbing the fluence estimate (optimally, the free-in-air fluence should be calculated with the phantom 'voided'), but the geometry divergences are likely not so radical that the ICRP data could not be expected to provide at least a ballpark estimate. In this case, the values published in ICRP 116 are $1.8 \times 10^{-13}$ Gy cm$^2$ and $2.2 \times 10^{-13}$ Gy cm$^2$ for ISO and ROT respectively, which are not very different from the value derived in the above simple test. Moreover, this quick analysis for just the male lungs could further be interpreted as providing some qualitative indication of the typical sizes of all the doses expected from the exercise, given: *a*) the penetrating nature of the photons, and hence the approximate dose uniformity that might be anticipated for the various organs within the body; and also *b*) the relatively small differences likely between male and female doses, again at least roughly. Certainly, the comparison could again be used to highlight as erroneous those organ doses that differed by one (or more) orders-of-magnitude from the other results within a given participant's dataset.

On a more rudimentary level, the initial implementation of the reference voxel phantoms within the Monte Carlo code, and the methods employed for calculating organ and effective doses, could also readily be verified. For example, participants could just remove the concrete floor from their model and change their source to an ideal anterior-posterior (AP) exposure (or ISO, ROT, posterior-anterior (PA), etc.), and then directly compare their results with the data tabulated in ICRP 116. This approach would not provide a check of the RBM values, because the recommended dose enhancement factors were not applied during the derivation of the published conversion coefficients, but would permit benchmarking of all other dose quantities. It is not known how many of the participants performed these sorts of 'sanity check' QA analyses on their models and data prior to submitting results for the current intercomparison exercise. But it may be remarked generally that ballpark benchmarking ought to be considered routine practice within all Monte Carlo modelling projects (voxel phantom or otherwise) to provide insight into the quality of results, and users need to be proactive in seeking innovative means to achieve it.

CONCLUSIONS

It is well-known that voxel phantom Monte Carlo simulations are complex. For example: input files can be large and unwieldy; software packages for geometry visualization may struggle to cope; CPU / RAM requirements may be at the limits of some PCs; and output data can be hard to process. It is also a non-trivial operation to manipulate the calculated organ doses and correctly determine effective dose from them, especially considering the additional complicating requirements of RBM (and endosteal) dosimetry. Moreover, without the availability of reference data or measured results, it is difficult to directly benchmark or check the output from non-standard exposure conditions, such as that featured in the current exercise. From this perspective, the benefits of intercomparison exercises are clear.

It is not possible to extrapolate specific conclusions on the current status of voxel phantom calculations within the global radiation protection community, just from this very limited sample set of thirteen participants. Indeed, those participants may themselves not necessarily even be representative: by its nature, an intercomparison exercise might plausibly garner more interest from individuals who consider themselves to be relatively inexperienced with the techniques, and are interpreting it as a training opportunity. Of course, that was one of its intended purposes.

Nevertheless, some general remarks might still be made. Firstly, the implementation of voxel phantom geometries can result in mistakes, and individuals may sometimes struggle to perform the calculations correctly. Following the ICRP 103 method for evaluating effective dose can also lead to difficulties, and even plausible-looking results may hide underlying inaccuracies in the contributory organ doses. Likewise, the recommended approach to bone marrow dosimetry can be poorly understood; this may also imply that accurate determination of endosteal tissue doses might be similarly problematic, though this has not been demonstrated explicitly in the current exercise. Following this observation, and also similar trends found in other exercises [Zankl *et al.*, 2021a], a separate article has been produced that describes the ICRP bone dosimetry method in more detail and provides further practical guidance for incorporating it into radiation transport codes [Zankl *et al.*, 2021b].

It must also be recalled that small statistical uncertainties outputted from Monte Carlo calculations may hide much larger systematic uncertainties, resulting perhaps from incorrect setting-up of the geometry and source or even post-processing mistakes in the analyses of results. In that regard, it may be commented that there is little gained by individuals striving for very high statistical precision in their modelling without also ensuring the actual accuracy of those results. To support that endeavour, proper QA should always be performed, with individuals continually asking themselves questions such as: '*do my results seem as expected*?'; '*are they self-consistent*?'; '*are they physically realistic, or obviously implausible*?'; '*what simple tests can I perform to check them*?'; etc. One subsidiary outcome from the current work might hence be a highlighting of the need for more training courses, such as on voxel phantoms, computational dosimetry, and even QA techniques themselves, which follows in turn to a suggestion that organizations such as EURADOS would naturally be well-placed to lead on such ventures. Finally, it is therefore obvious that intercomparison exercises of this type conducted by EURADOS WG6 are vital to the field of computational dosimetry. This conclusion is not just because they benefit the participating individuals themselves, nor even because they lead to the production and publication of reference geometries and solutions for non-standard exposures scenarios that can be of future help to novice users, but also because they are potentially useful for organizations such as ICRP in order to highlight any limitations in the general level of understanding within the community of some of their recommendations.

ACKOWLEDGEMENTS

J Eakins and C Huet acknowledge and thank all participants who contributed solutions to the intercomparison exercise, especially those not already listed as co-authors of this publication.

***Table 1.*** Atomic composition of the concrete by mass fraction.

| Element (Z) | 1 | 6 | 8 | 11 | 12 |
|---|---|---|---|---|---|
| Mass Fraction | 0.0221 | 0.002484 | 0.57493 | 0.015208 | 0.001266 |
| Element (Z) | 13 | 14 | 19 | 20 | 26 |
| Mass Fraction | 0.019953 | 0.304627 | 0.010045 | 0.042952 | 0.006435 |

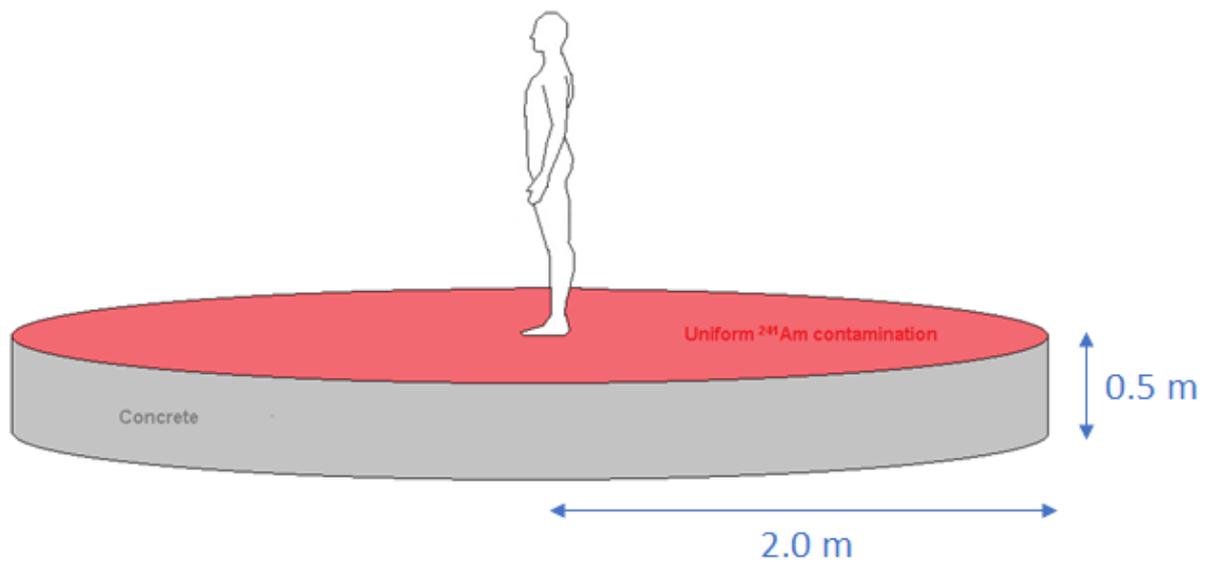

**Figure 1** The phantom surrounded by vacuum and standing on ground with a surface contamination of Am-241.

*Table 2.*   Codes, cross-sections and methods as reported by participants.

| Participant ID | Code | Kerma Approximation | Cross-Section Library | | RBM Method |
|---|---|---|---|---|---|
| | | | Photon | Electron | |
| 1 | FLUKA | No | EPDL97 | FLUKA model | ICRP 116 |
| 2 | FLUKA | No | EPDL97 | *Default* | ICRP 116 |
| 3 | VMC[a] | Yes | NIST XCOM | *Unspecified* | ICRP 116 |
| 4 | MCNP X2.7 | Yes | ENDF71 | *Unspecified* | ICRP 103 |
| 5 | MCNP X2.7 | Yes | *Default* | *Unspecified* | *Note*[e] |
| 6 | MCNP X2.7 | Yes / No[d] | MCPLIB04 | EL03 | ICRP 116 |
| 7 | MCNP 6.1.1 | Yes | MCPLIB84 | *Unspecified* | ICRP 116 |
| 8 | MCNP 6.1 | No | MCPLIB84 | EL03 | *Note*[f] |
| 9 | MCNP 6.2 | No | MCPLIB84 | EL03 | ICRP 116 |
| 10 | GEANT4[b] | No | EPDL97 | EEDL | *Note*[g] |
| 11 | GEANT4[c] | No | *Default* | *Default* | *Note*[h] |
| 12 | TRIPOLI-4 | No | ENDL97 | EEDL + Brem | 'Homemade' |
| 13 | *Unspecified* | *Unspecified* | *Unspecified* | *Unspecified* | *Unspecified* |
| *Organizer* | **MCNP X2.7** | **Yes** | **MCPLIB84** | **EL03** | **ICRP 116** |

[a] Visual Monte Carlo, version 09/18   [b] Build 10.05.p01   [c] Build 10.4
[d] Participant submitted two solutions : one with, and one without, the kerma approximation
[e] 'Dose response functions', further details unspecified
[f] 'Fluence-to-dose response function', further details unspecified
[g] 'Mass fraction correction to dose for each bone site', further details unspecified
[h] 'Fluence multiplied by dose response function', further details unspecified.

*Table 3.*   Reference organ and effective dose rates.

| Quantity | Dose rate (Gy s$^{-1}$ or Sv s$^{-1}$) | | | |
|---|---|---|---|---|
| | *Male* | | *Female* | |
| | Dose rate | std. unc. (%) | Dose rate | std. unc. (%) |
| Brain dose | $2.57 \times 10^{-8}$ | 0.28 | $3.09 \times 10^{-8}$ | 0.26 |
| Lung dose | $3.71 \times 10^{-8}$ | 0.17 | $5.36 \times 10^{-8}$ | 0.16 |
| Small intestine dose | $6.61 \times 10^{-8}$ | 0.15 | $6.87 \times 10^{-8}$ | 0.16 |
| Stomach dose | $5.11 \times 10^{-8}$ | 0.29 | $6.44 \times 10^{-8}$ | 0.27 |
| Red bone marrow dose | $3.73 \times 10^{-8}$ | 0.08 | $4.58 \times 10^{-8}$ | 0.08 |
| Effective dose | $6.29 \times 10^{-8}$   (0.1%) | | | |

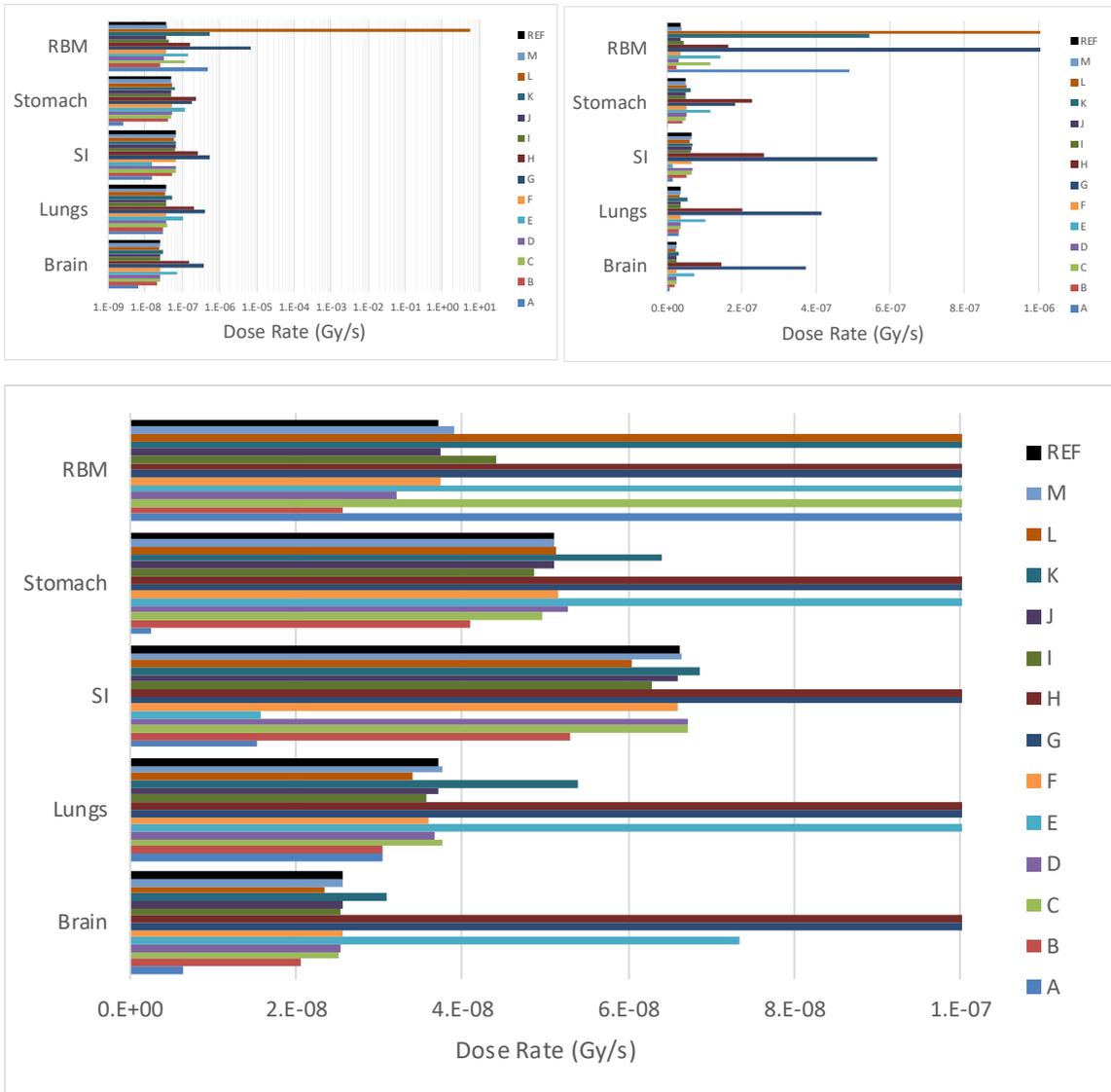

**Figure 2**   Organ dose rate data from the 13 participants for the male phantom, compared with reference data. (*Top left*) all data [Note: log scale]; (*Top right*) excluding extreme outliers; (*Bottom*) restricted range.

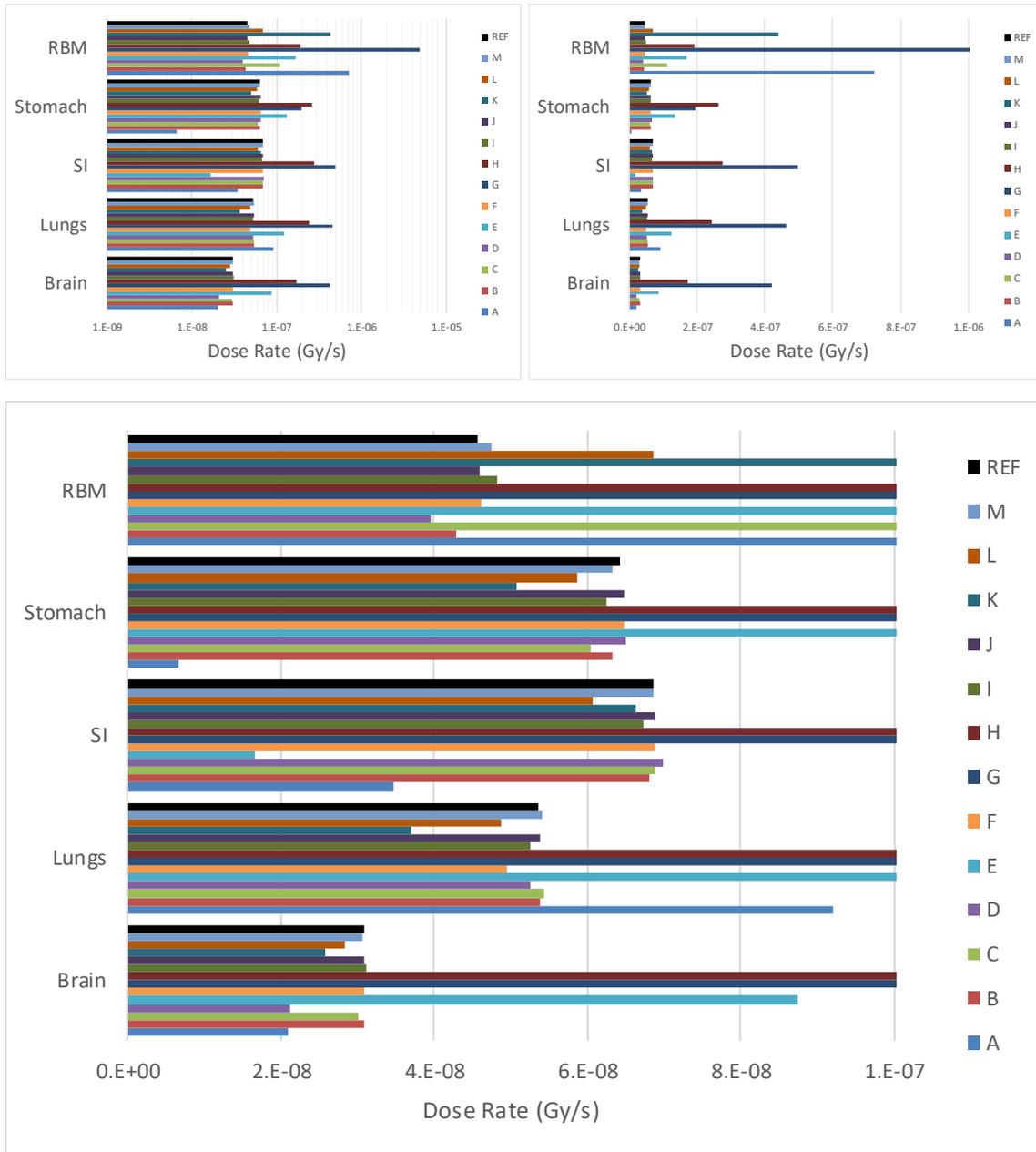

**Figure 3** Organ dose rate data from the 13 participants for the female phantom, compared with reference data. (*Top left*) all data [Note: log scale]; (*Top right*) excluding extreme outliers; (*Bottom*) restricted range.

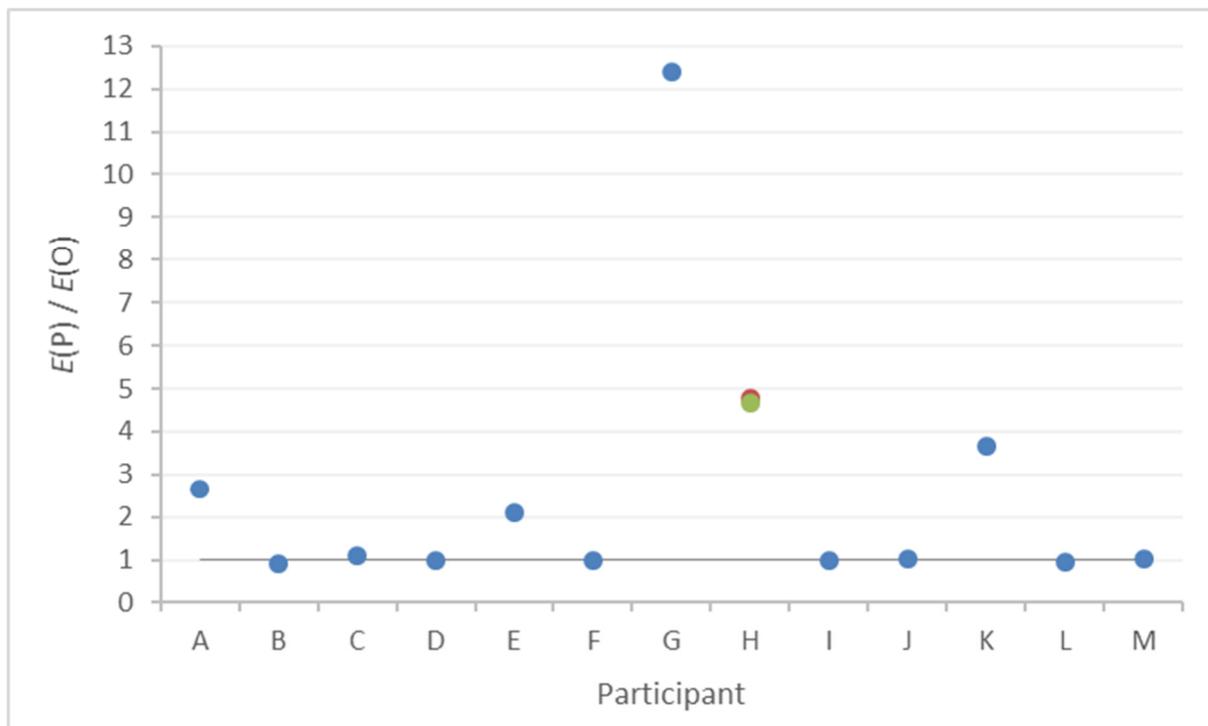

**Figure 4**   Ratio of effective dose rate data (*E*(P)) submitted by the participants to the reference effective dose rate (*E*(O)). Participant H erroneously provided results for the male (*green*) and female (*red*) phantoms separately.

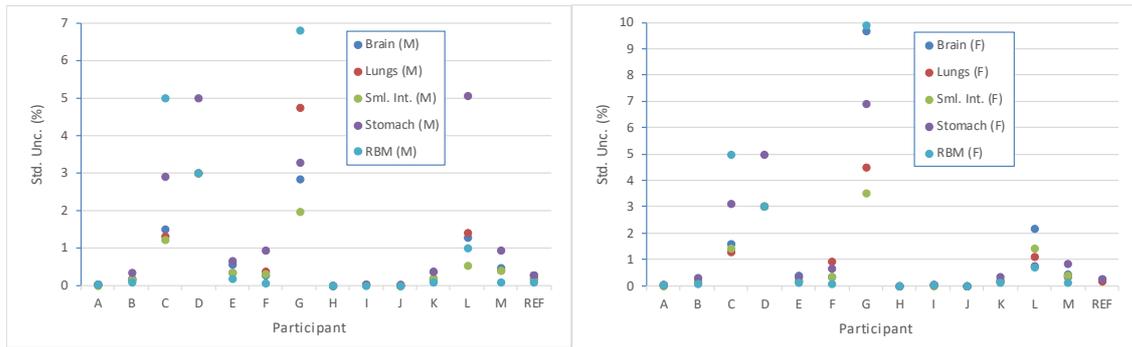

**Figure 5** Statistical standard uncertainties on the organ dose data reported by the participants and for the reference solutions: (*Left*) Male phantom; (*Right*) Female phantom.

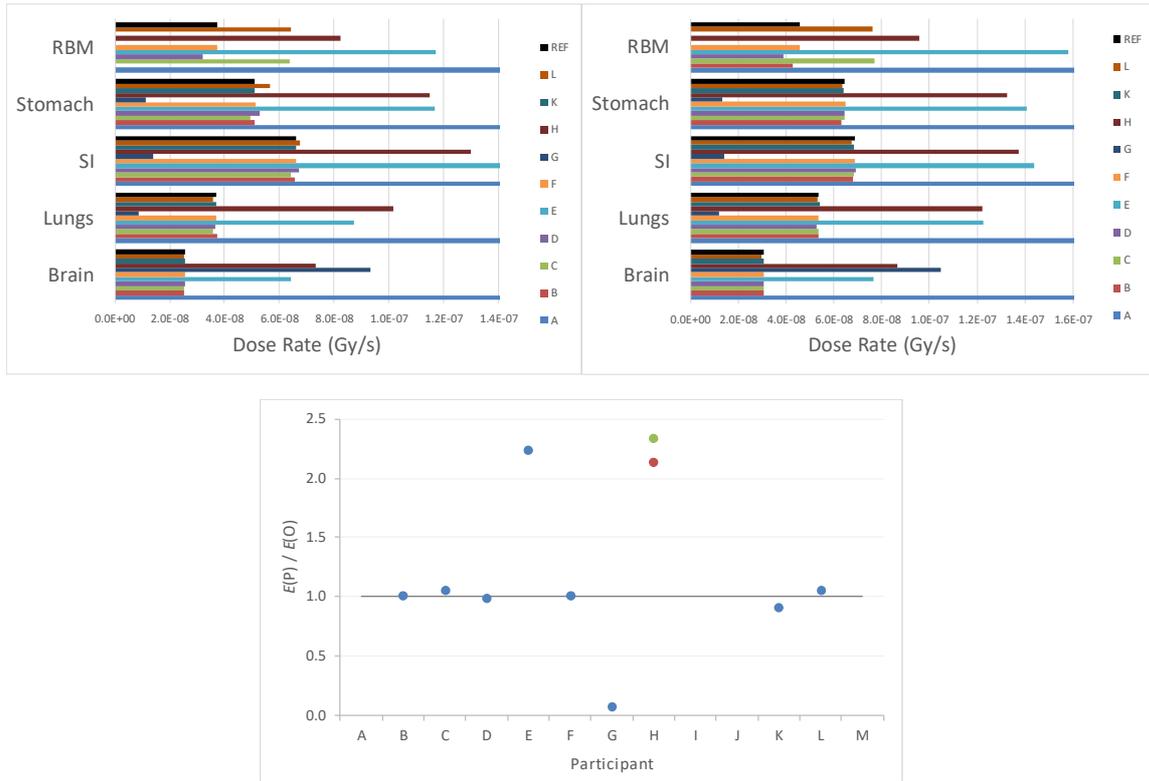

**Figure 6**  Revised submissions from 10 participants: (*Top left*) organ dose rates for the male phantom, (*Top right*) female phantom, and (*Bottom*) effective dose rate ratio *E*(P)/*E*(O). A restricted range has been applied to each plot.

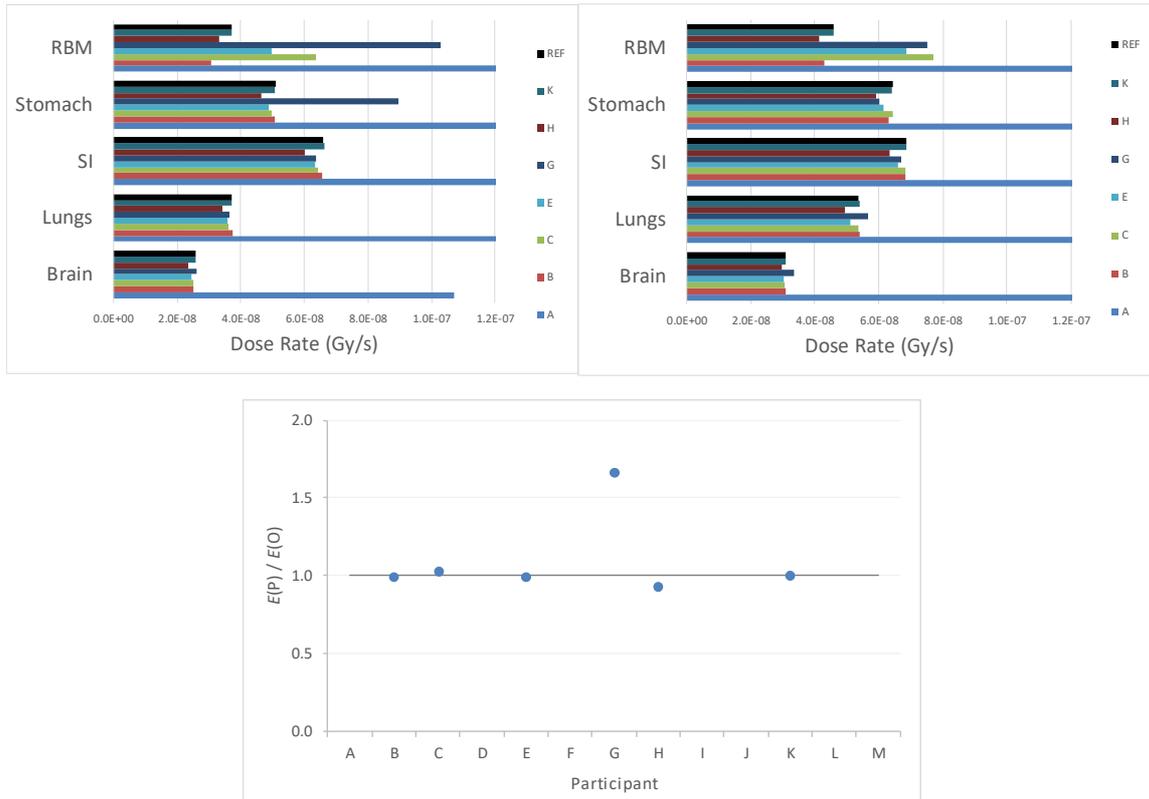

**Figure 7**  Second set of revised submissions from 7 participants: (*Top left*) organ dose rates for the male phantom, (*Top right*) female phantom, and (*Bottom*) effective dose rate ratio *E*(P)/*E*(O). A restricted range has been applied to each plot.